\documentclass[fleqn,usenatbib]{mnras}

\usepackage{newtxtext,newtxmath}

\usepackage[T1]{fontenc}

\DeclareRobustCommand{\VAN}[3]{#2}
\let\VANthebibliography\thebibliography
\def\thebibliography{\DeclareRobustCommand{\VAN}[3]{##3}\VANthebibliography}

\usepackage{graphicx}	
\usepackage{booktabs}
\usepackage[flushleft,para]{threeparttable}
\usepackage{amsmath}	
\usepackage{siunitx}
\usepackage{balance}
\usepackage{hyperref}
    \hypersetup{colorlinks=true,linkcolor=blue,citecolor=blue,urlcolor=blue}
\usepackage[nowarn]{glossaries}
    \makeglossaries

\newacronym{ARVE}{ARVE}{Analyzing Radial Velocity Elements}

\newacronym{BIS}{BIS}{bisector inverse slope}
\newacronym{BJD}{BJD}{Barycentric Julian Date}

\newacronym{CB}{CB}{convective blueshift}
\newacronym{CCF}{CCF}{cross-correlation function}

\newacronym{DACE}{DACE}{Data \& Analysis Center for Exoplanets}
\newacronym{DRS}{DRS}{Data Reduction Software}

\newacronym{EPRV}{EPRV}{Extreme Precision RV}

\newacronym{FAP}{FAP}{false alarm probability}
\newacronym{FWHM}{FWHM}{full-width at half-maximum}

\newacronym{GLS}{GLS}{generalised Lomb-Scargle}
\newacronym{GP}{GP}{Gaussian process}

\newacronym{HARPS-N}{HARPS-N}{High Accuracy Radial velocity Planet Searcher for the Northern hemisphere}
\newacronym{HMI}{HMI}{Helioseismic and Magnetic Imager}

\newacronym{JSOC}{JSOC}{Joint Science Operations Center}

\newacronym{LBL}{LBL}{line-by-line}

\newacronym{MCMC}{MCMC}{Markov Chain Monte Carlo}
\newacronym{MLR}{MLR}{multi-linear regression}
\newacronym{MS}{MS}{main-sequence}

\newacronym{PBP}{PBP}{part-by-part}
\newacronym{PCA}{PCA}{principal component analysis}

\newacronym{QP}{QP}{quasi-periodic}

\newacronym{RMS}{RMS}{root mean square}
\newacronym{RV}{RV}{radial velocity}

\newacronym{S/N}{S/N}{signal-to-noise ratio}
\newacronym{SDO}{SDO}{Solar Dynamics Observatory}
\newacronym{SME}{SME}{Spectroscopy Made Easy}

\newacronym{TNG}{TNG}{Telescopio Nazionale Galileo}

\newacronym{VALD}{VALD}{Vienna Atomic Line Database}
\usepackage{orcidlink}
\usepackage[normalem]{ulem}

\title[GP regression of temperature-dependent RVs]{Gaussian process regression of temperature-dependent radial velocities}

\author[F. Rescigno and K. Al Moulla]{
Federica Rescigno$^{\orcidlink{0000-0002-0594-7805}\,1,2}$\thanks{E-mail: f.rescigno@bham.ac.uk} and
Khaled Al Moulla$^{\orcidlink{0000-0002-3212-5778}\,3}$
\\
$^{1}$Department of Astrophysics, University of Birmingham, Edgbaston, Birmingham, B15 2TT, UK\\
$^{2}$Department of Astrophysics, University of Exeter, Stocker Rd, Exeter, EX4 4QL, UK\\
$^{3}$Observatoire Astronomique de l'Université de Genève, Chemin Pegasi 51, 1290 Versoix, Switzerland\\
}

\date{Accepted XXX. Received YYY; in original form ZZZ}

\pubyear{2024}

\begin{document}
\label{firstpage}
\pagerange{\pageref{firstpage}--\pageref{lastpage}}
\maketitle

\begin{abstract}
Gaussian processes (GPs) described by quasi-periodic covariance functions have in recent years become a widely used tool to model the impact of stellar activity on radial velocity (RV) measurements. We perform a GP regression analysis on solar RV time series measured from spectral segments formed at different temperatures within the photosphere in order to evaluate the relation between the best-fit GP kernel hyperparameters and the observed activity signal as a function of temperature. The posterior distributions of the hyperparameters show subtle differences between high- and low-activity phases and as a function of the spectral formation temperature range, which could have implications on the characteristics of the activity signal and its optimal modelling. For the temperature-dependent RVs, we find that at high and low activity alike, the minimal RV dispersion is obtained at intermediately cool temperature ranges ($\SI{4000}{}{-}\SI{4750}{K}$), for both the observed and GP model-subtracted RVs. Finally, we compare and correlate our temperature-dependent RVs with RV components derived from disk-resolved Dopplergrams of the Sun, for which we find a consistently strong correlation between RVs related to hotter temperature ranges and the dominant RV component due to the inhibition of convection.
\end{abstract}




\section{Introduction}\label{Sect:1}

In the last three decades, \acrfull*{RV} analyses have successfully detected more than 1,000 different exoplanets and have contributed to characterising the masses of thousands more, from gas giants to rocky planets. As the precision and stability of spectrographs approach the centimetre-per-second realm, the signals of Earth-like exoplanets in Earth-like orbits are coming within reach of detection. Nevertheless, the greatest challenge for the precise and accurate characterisation of exoplanetary signals in the \acrfull*{EPRV} regime remains stellar variability \citep{Meunier+2010,Dumusque+2011,Fischer+2016,Crass+2021,Meunier2021}. RV variations generated by the multitude of physical processes shaping the stellar photo- and chromosphere can mimic or obscure the presence of planets leading to false detections, non-detections or inaccurate mass measurements \citep[e.g.,][]{Rajpaul+2016,John+2022}. It is therefore of vital importance to understand these stellar activity signals in order to average or model them out.

Stellar activity indicators derived either from the observed spectra, e.g., the $S$-index \citep{Wilson1968}, or from their \acrlongpl*{CCF} \citep[CCFs;][]{Baranne+1996,Pepe+2002}, e.g., the \acrlong*{BIS} \citep[BIS;][]{Queloz+2001} or the \acrlong*{FWHM} \citep[FWHM;][]{Queloz+2009}, have historically been utilised to discriminate stellar variability from planetary signals \citep[e.g.,][]{Desort+2007,Rajpaul+2015,Costes+2021}. These proxies are not sensitive to the Doppler-shift generated by the gravitational pull induced by a planet, and can therefore be used as comparative data sets. However, recent works have demonstrated that traditional activity indicators are not sufficient for a full mitigation of the stellar variability \cite[e.g.,][]{Haywood+2016,Cretignier+2024} and often even fail to consistently measure the stellar rotational period, a vital basic characteristic of some of these signals \citep{Nava+2022, Rescigno+2024b}. \cite{AlMoulla+2024} instead proposed the use of RV measurements extracted from various line-forming temperature ranges in the spectra as a stellar variability proxy for activity mitigation. They showed that detrending the full-spectrum RVs with both the $S$ index and these temperature-dependent RVs reaches lower scatter than by correcting the RVs with the $S$ index alone. This implies that RVs derived from spectral line segments formed at various temperatures exhibit different sensitivity to activity, and could therefore improve our understanding of it and allow us to develop better mitigation methods.

The stellar variability present in these temperature-dependent RVs can then be modelled with Gaussian processes (GPs). GP regression coupled with \acrfull*{MCMC} parameter space exploration algorithms \citep[e.g.,][]{Foreman-Mackey+2017} is in fact commonly implemented in current analyses to predict signals generated by stellar activity \citep[e.g.,][]{Rajpaul+2015,Serrano+2018,Barros+2020,Dalal+2024}. GPs not only allow us to flexibly model the RV and indicator time series without assuming a deterministic form, but they can also be a powerful tool to investigate the evolution of their covariance. In practical terms, by studying the changes of the kernel hyperparameters that best fit each different set of data we can learn about the evolution of the stellar variability in time.

In this work, in order to better understand how stellar variability affects the RVs extracted from lines formed at different temperatures, we compare the functional form of the covariance of temperature-dependent RV time series derived following the method of \cite{AlMoulla+2022}. We investigate the relationship between the best-fit kernel hyperparameters obtained via GP regression, the 11 temperature ranges, and two time intervals (at high and low activity levels), for a total of 22 unique time series. Sun-as-a-star spectral data (meaning observational data in which the Sun is observed as a point-like light source) have been selected for this work, as the Sun is the ideal testing ground for new mitigation techniques, offering both unparalleled cadence and baseline. This paper is structured as follows. The solar spectral data is presented in Sect.~\ref{Sect:2}. The method employed to extract the RV measurements is described in Sect.~\ref{Sect:3.1}. We cover the GP modelling technique and the choice of kernel formulation in Sect.~\ref{Sect:3.2}. The results of this analysis are then presented in Sect.~\ref{Sect:4}. We conclude in Sect.~\ref{Sect:5}.

\section{Data}\label{Sect:2}

We analyse the first three years of solar data \citep{Dumusque+2021} from the \acrlong*{HARPS-N} \citep[HARPS-N;][]{Cosentino+2012,Cosentino+2014} high-resolution spectrograph, mounted on the \acrfull*{TNG} at the Roque de los Muchachos Observatory in La Palma, Spain, taken from 18 July 2015 to 16 July 2018, i.e., \acrfull*{BJD} 2,457,222 to 2,458,316.

The HARPS-N solar telescope \citep{Dumusque+2015,Phillips+2016,CollierCameron+2019} is comprised of a $\SI{7.6}{\centi\meter}$ achromatic lens. The sunlight is then sent to the HARPS-N spectrograph after being scrambled by an integrating sphere to mimic a point-like source. The telescope observes the Sun at a cadence of $\SI{5.5}{min}$ during day-time. The spectra are heliocentrically corrected by removing the Doppler reflex motion of all Solar System planets, and thereafter processed with \texttt{YARARA} \citep{Cretignier+2021,Cretignier+2023}, which is a post-processing software correcting for instrumental systematics, telluric contamination and stellar activity in multiple sub-routines. Since we are interested in studying the characteristics of stellar activity, we re-inject the solar activity correction at the spectral level. In order to maximise the efficacy of these corrections, \texttt{YARARA} bins the data over each day, leaving us with 655 spectra.

Due to the computational expense of GP regression, we focus our analysis on two temporal subsets of the solar data during which the Sun exhibited higher and lower levels of RV variability, respectively. In order to better constrain the GP models, we choose time intervals of 140 days in length. This baseline allows us to cover multiple solar rotations as well as probe the evolution timescale of active regions on the surface of the Sun. The high activity interval is selected from 5 June to 23 October 2016, i.e., BJD 2,457,544.5 to 2,457,684.5, as indicated by the orange-shaded region in Fig.~\ref{Fig:01}. This time interval does not include the maximum solar activity level recorded by the HARPS-N data, as we choose to instead prioritise the completeness of the time series over the chosen time range in order to minimise the impact of uneven cadence due to lack of data caused by bad weather. The level of solar activity in the selected time interval is nevertheless comparable to that of the start of the HARPS-N dataset (closer in time to the maximum of Cycle 24) and remains significantly higher than that of the other selected section. The low activity interval is selected from 20 February to 10 July 2018, i.e., BJD 2,458,169.5 to 2,458,309.5, shown in Fig.~\ref{Fig:01} as a green-shaded region. Both time intervals include $\sim$100 daily-binned RVs, with some gaps due to weather conditions.

\section{Methods}\label{Sect:3}

\subsection{RV extraction}\label{Sect:3.1}

Temperature-dependent RV extraction is performed with the \acrlong*{ARVE} (\texttt{ARVE}; Al Moulla in prep.) software. \texttt{ARVE}\footnote{\url{https://github.com/almoulla/arve}} is an open-source Python package enabling multi-functional extraction of stellar RVs, e.g., with built-in CCF binary masks or \acrfull*{LBL} template-matching \citep{Anglada-Escude&Butler2012,Dumusque2018,Artigau+2022}.

RVs can also be extracted as a function of the spectral average formation temperature, $T_{1/2}$, defined as the photospheric temperature at which the cumulative flux contribution function reaches 50\% of its maximum value \citep[see][]{AlMoulla+2022}. By retrieving a formation temperature map from a spectral synthesis model, one can template-match the flux spectral segments corresponding to a certain formation temperature range against a high \acrfull*{S/N} master spectrum in order to obtain RVs at various photospheric depths.

\texttt{ARVE} extracts temperature-dependent RVs using pre-computed formation temperature maps for a grid of spectral types. The spectral syntheses are performed with \texttt{PySME} \citep{Wehrhahn+2023}, a Python implementation of \acrlong*{SME} \cite[\texttt{SME};][]{Valenti&Piskunov1996,Piskunov&Valenti2017}, which uses \texttt{MARCS} model atmospheres \citep{Gustafsson+2008} and line lists from the \acrlong*{VALD}\footnote{\url{http://vald.astro.uu.se}} \citep[VALD;][]{Piskunov+1995,Kupka+2000,Ryabchikova+2015}. For the Sun, we adopt the default \texttt{PySME} parameters for effective temperature, $T_\mathrm{eff}\,{=}\,\SI{5770}{K}$, surface gravity, $\log{g}\,{=}\,4.40$, and metallicity, $[\mathrm{Fe}/\mathrm{H}]\,{=}\,0.00$, and elemental abundances from \cite{Asplund+2009}. Despite the irrelevance of spectral broadening to the derivation of formation temperatures, in the subsequent steps we exclude poorly synthesised spectral lines because their inferred formation temperatures would be unreliable. We aim to model the solar flux spectrum as closely as possible to observations. The spectral synthesis is therefore broadened with rotational broadening, $v\sin{i}\,{=}\,\SI{1.63}{\kilo\meter\per\second}$, micro- and macroturbulences, $v_{\mathrm{mic}}\,{=}\,\SI{0.85}{\kilo\meter\per\second}$ and $v_{\mathrm{mac}}\,{=}\,\SI{3.98}{\kilo\meter\per\second}$, respectively, adopted from \cite{Valenti&Fischer2005}, and is convolved with a Gaussian instrumental profile corresponding to the $R\,{=}\,115,000$ spectral resolution of HARPS-N.

The solar spectral line formation temperature map for the HARPS-N wavelength coverage ($\SI{390}{}{-}\SI{690}{\nano\meter}$) ranges from $4094$ to $\SI{5501}{K}$ \citep[see Fig.~2 in][]{AlMoulla+2022}. Variations across small wavelength ranges (a few Ångströms) originate from individual line opacities, yielding cooler formation temperatures in line cores and higher formation temperatures in line wings. Variations across large wavelength ranges (a few hundreds of Ångströms) are due to continuous absorption, which in the optical is dominated by the negative hydrogen ion, yielding generally cooler formation temperatures at redder wavelengths up until ${\sim}\SI{1}{\micro\meter}$ \citep{Gray2008}.

\subsection{GP regression modelling}\label{Sect:3.2}

GPs allow us to model a time series without making assumptions about its deterministic form, and instead rely on an understanding of the shape of its covariance. In this work, we assume that the solar RV time series is dominated by variability generated by the photometric inhomogeneities and by the suppression of convective blueshift caused by the presence of dark spots and bright faculae. We thus expect the temperature-dependent RVs to also be sensitive to the same rotationally-modulated effects. We can therefore model the data with a \acrfull*{QP} kernel, as described in \cite{Haywood+2014}, with a supplementary white noise component in the form:
\begin{equation}
    k(t_i,t_j) = A^2 \exp\left[ -\frac{|t_i - t_j|^2}{\tau^2} - \frac{\sin^2 \left(\frac{\pi |t_i - t_j|}{P_{\rm rot}} \right)}{\mu^2} \right] + \delta_{i,j}\beta^2.
    \label{Eq:01}
\end{equation}
In this equation, $k(t_i,t_j)$ represents the covariance function of the quasi-periodic variation between time steps $t_i$ and $t_j$. The QP kernel requires four hyperparameters: $A$ is the amplitude of the variation. $\tau$ describes the timescale of the evolution of the quasi-periodicity of the data, and it is most often referred to as the evolution timescale of the active regions producing the signal. $P_{\mathrm{rot}}$ is the period of the variation, in this case the stellar rotation. $\mu$ is a ``smoothness'' factor, and defines the harmonic complexity of the fit (larger $\mu$ values yield a ``smoother'' fit in-between each period). For ease of interpretation, from hereon we will refer to $\mu$ as the inverse harmonic complexity variable. The smaller the value of $\mu$, the more strongly correlated are data points separated in time by multiples of the period $P_{\mathrm{rot}}$ than data points separated by other time lags. Finally, $\beta$ serves as a jitter term, and it represents the contribution of white noise in the data generated from their inherent precision; $\delta_{i,j}$ is the Kronecker-Delta which allocates the jitter to the diagonal elements of the covariance matrix. This jitter term can, in some cases, also be inflated significantly above the uncertainty level of the data for time series where the chosen covariance form does not successfully describe all types of variability present in the data.

In this analysis, the GP regression is accomplished with \texttt{MAGPy\_RV}\footnote{\url{https://github.com/frescigno/magpy_rv}} \citep{Rescigno+2023,Rescigno+2024}. The best-fit kernel hyperparameters are determined through an iterative process with an affine invariant MCMC algorithm and logarithmic likelihood optimisation. To expedite the process, we parallelise the code across multiple computing cores. In all of our runs, we evolve 100 chains for 50,000 iterations each, and discard a starting burn-in phase of 10,000 steps (equal to 20\% of the total iterations). The health of the chains and their convergence are assessed via Gelman-Rubin statistic.

For the sake of comparability, we apply the same priors to the GP regression of all temperature-dependent RV time series, as well as for both the time ranges chosen over different activity levels. The choice of priors is detailed in Table~\ref{Tab:01}. The amplitude is bound with a wide uniform prior between 0 and $\SI{20}{\meter\per\second}$. We require $\tau$ to be constrained between 0 and 140 days, which is the baseline of the time series considered. We bind the rotational period $P_{\mathrm{rot}}$ with a wide Gaussian prior centred around the Carrington solar rotation period of 27 days with standard deviation of 10 days, derived from preliminary Fourier analysis of the data. The inverse harmonic complexity is allowed to vary between 0 and 1 with a uniform prior. When interpreting the posteriors of $\mu$ we however need to be cautious: very small or very large inverse harmonic complexity values usually indicate overfitting of the white (or unmodelled) noise as part of the GP harmonic complexity, or that all inner-period signals generated as harmonic components of the stellar rotation by complex active region distributions are instead assigned to the jitter term, respectively. Finally, we bind the jitter term with a Gaussian prior centred around the mean uncertainty of the RV time series considered, and with a standard deviation equal to $\SI{0.2}{\meter\per\second}$. We note that this last prior is the only one which includes a dependency on the selected data set---as the uncertainties in the derived RVs vary due to the spectral fraction used in the extraction, so will the centre of this Gaussian prior. This is done in order to properly represent the amount of white noise in each time series.

\begin{table}
\centering
\caption{Solar activity priors applied to the GP regression analyses described in Sect.~\ref{Sect:3.2}. Priors have been identified as follows: $\mathcal{U}$ is a uniform prior with a minimum and maximum value, $\mathcal{N}$ is a Gaussian prior with a central value and standard deviation.}
\begin{tabular*}{\linewidth}{l @{\extracolsep{\fill}} l @{\extracolsep{\fill}} l @{\extracolsep{\fill}} l}
\toprule
\midrule
Parameter                   & Symbol             & Unit                       & Prior                                                     \\
\midrule
Amplitude                   & $A$                & $\SI{}{\meter\per\second}$ & $\mathcal{U}[0,20]$                                       \\
Timescale                   & $\tau$             & days                       & $\mathcal{U}[0,140]$                                      \\
Period                      & $P_{\mathrm{rot}}$ & days                       & $\mathcal{N}[27,10]$                                      \\
Inverse harmonic complexity & $\mu$              &                            & $\mathcal{U}[0,1]$                                        \\
Jitter                      & $\beta$            & $\SI{}{\meter\per\second}$ & $\mathcal{N}[\overline{\sigma_{\mathrm{RV}}}^{\dag},0.2]$ \\
\bottomrule
\end{tabular*}
\begin{tablenotes}
\small
\item $^{\dag}$ This value is equal to the mean value of the uncertainties of all extracted RVs for the considered temperature range.
\end{tablenotes}
\label{Tab:01}
\end{table}

\begin{figure*}
	\includegraphics[width=\linewidth]{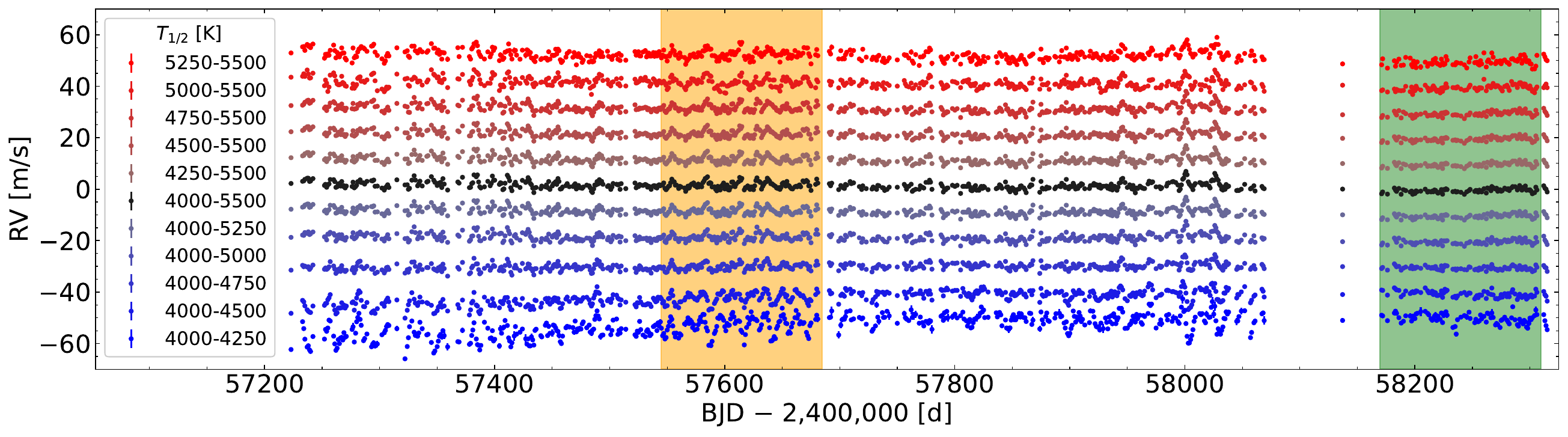}
    \caption{Temperature-dependent solar RVs as a function of BJD. The RV time series are computed using spectral segments formed at different average formation temperatures, $T_{1/2}$. The black points represent the RVs computed using roughly the entire formation temperature range ($\SI{4000}{}{-}\SI{5500}{K}$), the blue points represent RVs computed at sequentially cooler temperature regimes by decreasing the upper bound, and the red points represent RVs computed at sequentially hotter temperature regimes by increasing the lower bound. The orange- and green-shaded areas indicate high- and low-activity 140-day time intervals, respectively, on which we apply the GP regression. The RV curves have been artificially offset in increments of $\SI{10}{\meter\per\second}$ for visual purposes. The plotted errorbars are smaller than the markers.}
    \label{Fig:01}
\end{figure*}

\section{Results}\label{Sect:4}

\subsection{Temperature-dependent RV time series}\label{Sect:4.1}

In this work, we intend to study the transitional behaviours of temperature-dependent RVs and the best-fit hyperparameters of their corresponding GP model. We therefore opt to select our formation temperature ranges to be sequentially hotter or cooler, whilst keeping one boundary fixed. This approach is subtly different from the approach in \cite{AlMoulla+2022}, where they selected sequentially smaller but disjoint temperature ranges. As a reference case, we compute temperature-dependent RVs for the entire available temperature range (see Sect.~\ref{Sect:3.1}), rounding the bounds to $\SI{4000}{}{-}\SI{5500}{K}$ for simplicity. We thereafter compute RV time series for sequentially cooler temperature ranges, where the upper bound decreases in steps of $\SI{250}{K}$ until the remaining range spans $\SI{250}{K}$, i.e., $\SI{4000}{}{-}\SI{4250}{K}$. Likewise, we compute RV time series for sequentially hotter temperature ranges, where the lower bound increases in steps of $\SI{250}{K}$ until the remaining range spans $\SI{250}{K}$ as well, i.e., $\SI{5250}{}{-}\SI{5500}{K}$. We therefore extract a total of 11 RV time series, five with sequentially cooler temperature ranges, five with sequentially hotter temperature ranges, and one that includes the entire temperature range.

The resulting RV time series are shown in Fig.~\ref{Fig:01}, colour-coded based on the temperature range used in the extraction. From visual inspection, it becomes clear that the stellar activity signal is not constant throughout the line-forming temperature ranges. As a general trend, the RV peak-to-peak variations increase toward the temperature extremes. Part of this amplified scatter can be attributed to the increased noise, as a lesser fraction of the spectra is used to compute the RVs for the smaller temperature ranges. However, most of the dispersion is caused by depth-evolving stellar variability, since the observed variations greatly exceed the RV uncertainties computed from the intrinsic photon noise.

\begin{figure*}
	\includegraphics[width=\linewidth]{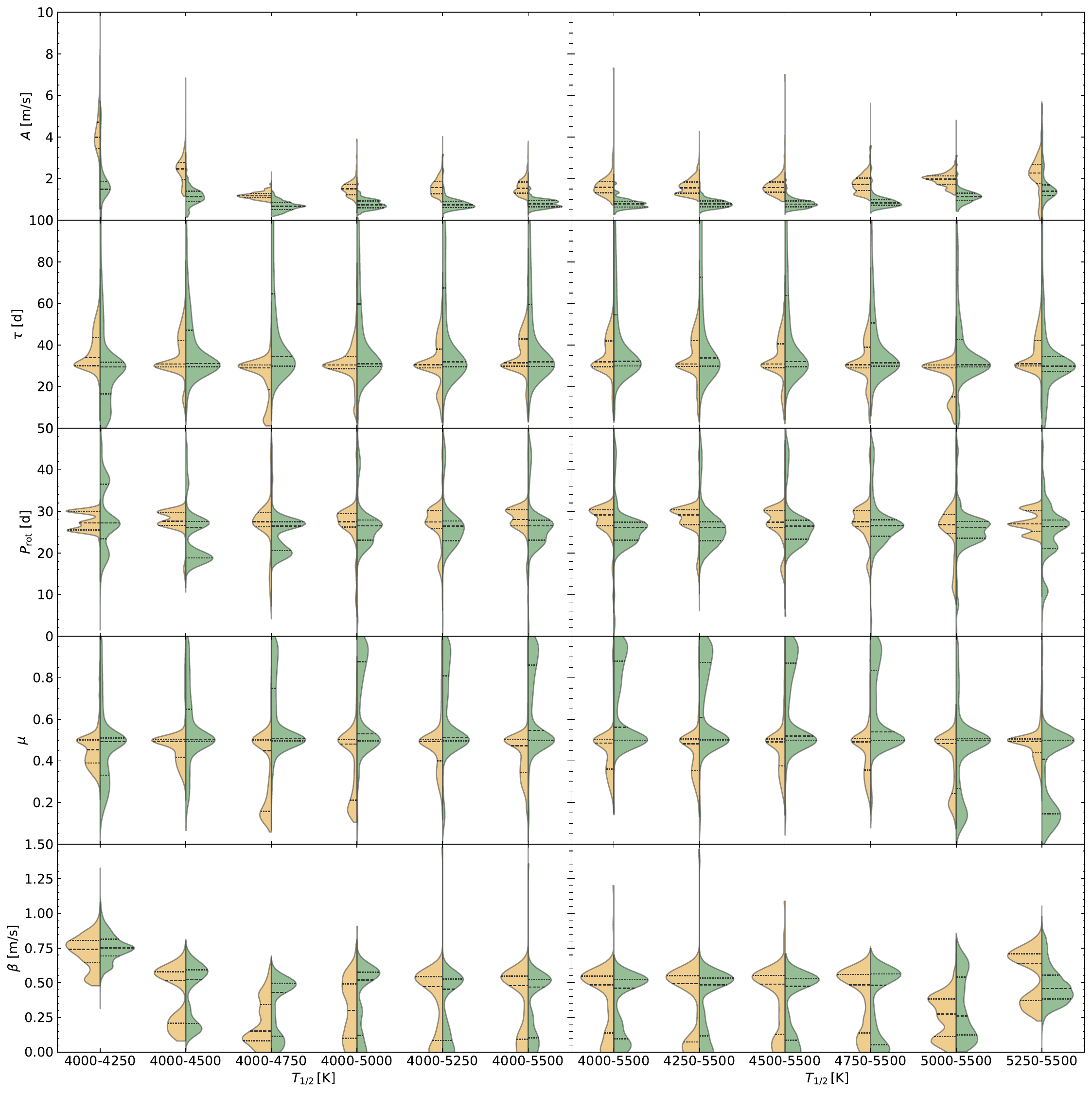}
    \caption{Violin plots of the posterior distributions of the kernel hyperparameters from Eq.~\ref{Eq:01} after GP regression. \textit{Left panels}: GP hyperparameters for the sequentially cooler temperature ranges, for both the high-activity (orange) and low-activity (green) time intervals indicated in Fig.~\ref{Fig:01}. The large-dashed lines show the $50^{\mathrm{th}}$ percentile in each distribution, and the small-dashed lines show the $16^{\mathrm{th}}$ and $84^{\mathrm{th}}$ percentiles, i.e., ${\pm}1\sigma$ intervals. \textit{Right panels}: Same as the left panels, but for the sequentially hotter temperature ranges.}
    \label{Fig:02}
\end{figure*}

\begin{figure*}
	\includegraphics[width=\linewidth]{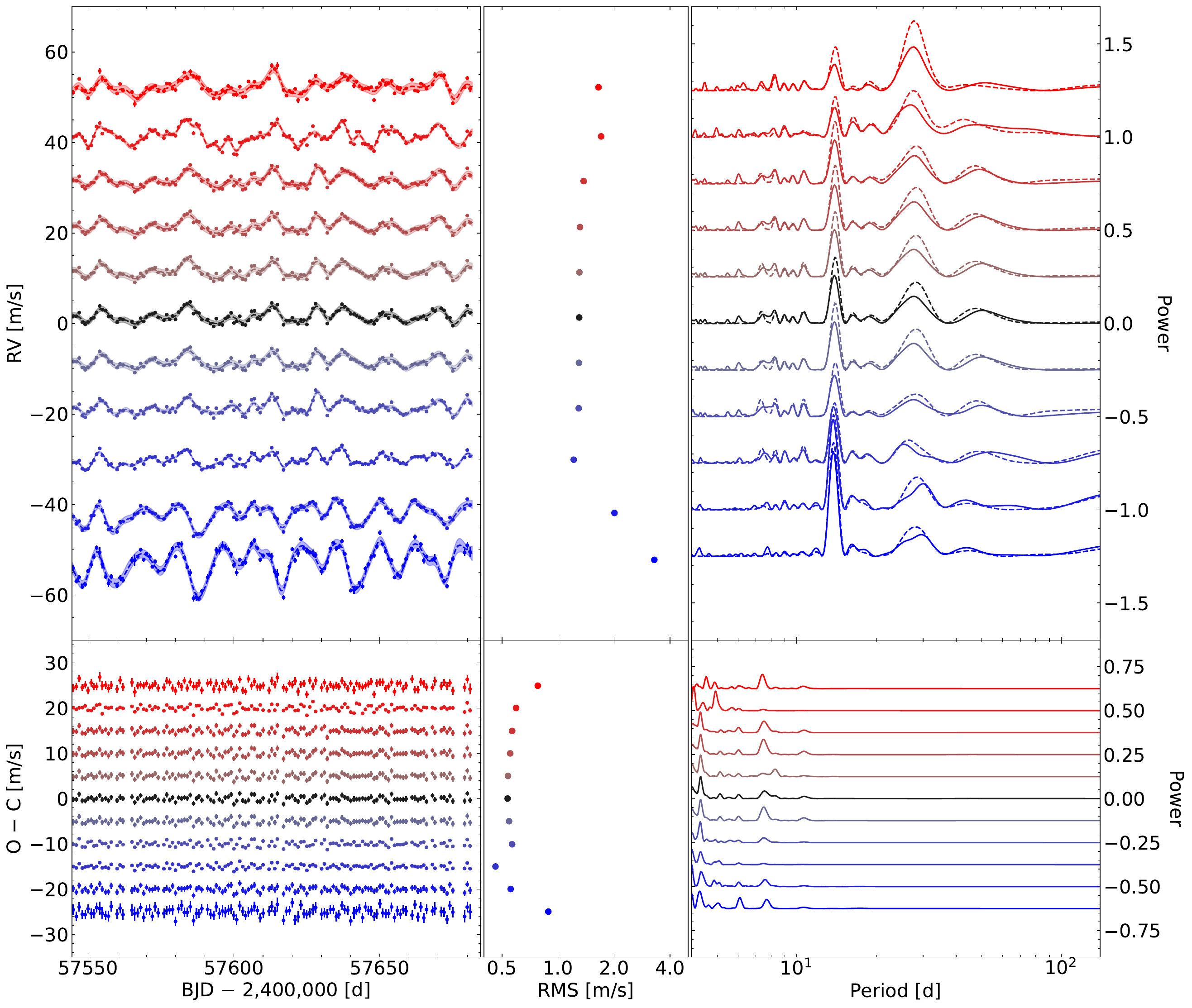}
    \caption{GP regression of the high-activity time interval. \textit{Top left panel}: RV time series at different formation temperature ranges. The curve colours represent the same average formation temperatures as in Fig.~\ref{Fig:01} Observations are shown as points with errorbars, and the best-fit GP models are shown as dashed lines with shaded $1\sigma$-intervals. \textit{Top centre panel}: RV RMS for the data in the top left panel. Note the logarithmic $x$-axis. \textit{Top right panel}: GLS periodograms of the observations (solid lines) and GPs (dashed lines) from the top left panel. \textit{Bottom panels}: Same as top panels but for the RV residuals obtained by subtracting the GP models from the observations.}
    \label{Fig:03}
\end{figure*}

\begin{figure*}
	\includegraphics[width=\linewidth]{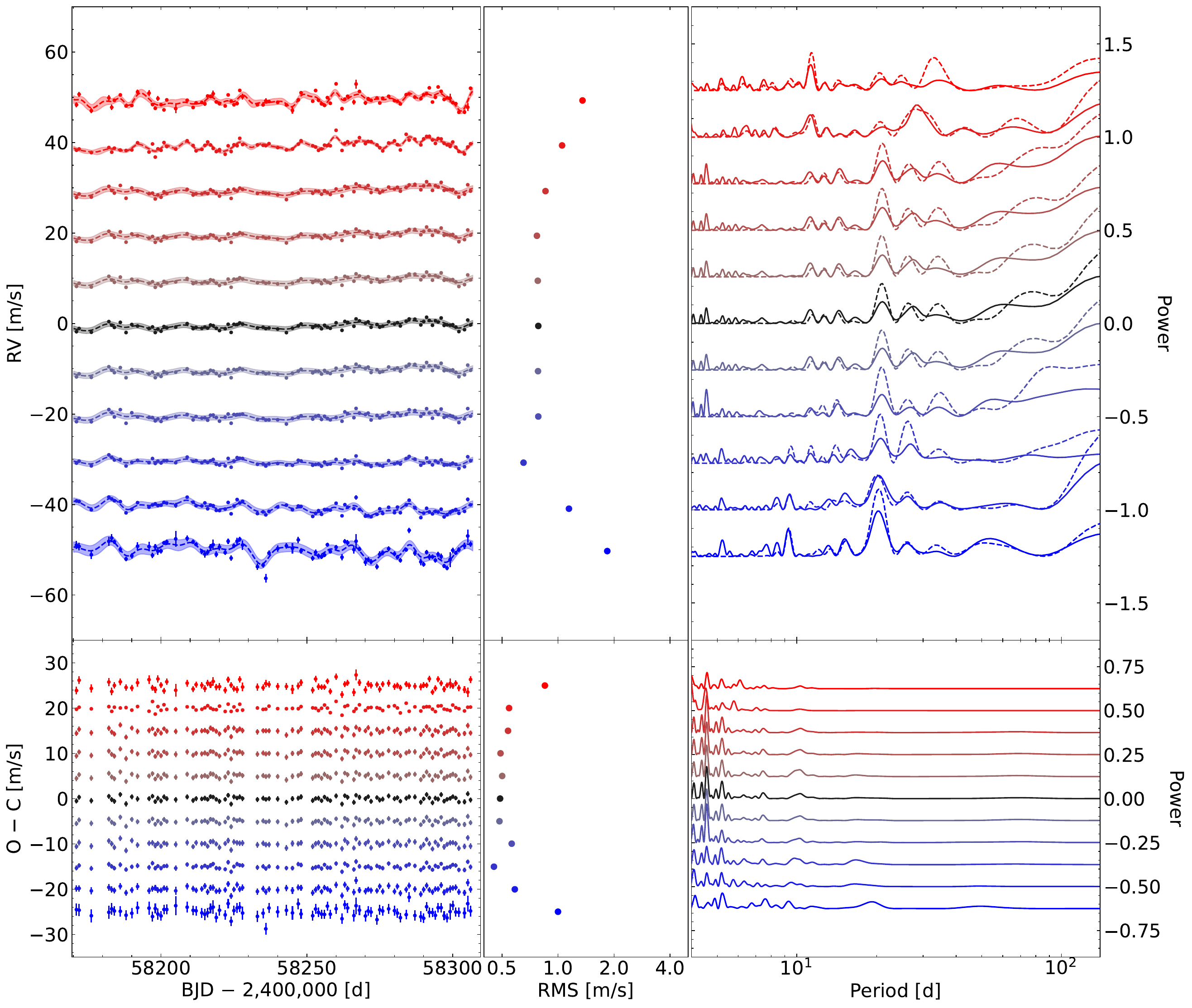}
    \caption{Same as Fig.~\ref{Fig:03}, but for the low-activity time interval.}
    \label{Fig:04}
\end{figure*}

\subsection{Covariance function hyperparameters}\label{Sect:4.2}

The posterior distributions produced via MCMC optimisation for all hyperparameters used to describe the GP covariance are plotted in Fig.~\ref{Fig:02}. The results of the high-activity time interval are shown in orange, while those of the low-activity time interval are plotted in green. Each formation-temperature range is identified by its temperature bounds, going from the coolest bin on the left to the hottest on the right. The full spectral range ($\SI{4000}{}{-}\SI{5500}{K}$) is analysed twice independently, as a further validation. All posteriors are computed with the same total number of iterations and are therefore directly comparable. All posteriors do not include results from the burn-in phase. The large-dashed lines within each distribution represent the $50^{\mathrm{th}}$ percentile values of each posterior, while the small-dashed lines indicate the $16^{\mathrm{th}}$ and $84^{\mathrm{th}}$ percentiles, respectively. In single-peaked distributions they straightforwardly represent the best-fit hyperparameter value and its ${\pm}1\sigma$ uncertainties. We however recommend caution when analysing more complex posterior distributions.

The amplitude, $A$, of the covariance reflects well the behaviour of the data. For all temperature ranges, the posterior distributions for high-activity time series peak at larger $A$ values than their corresponding low-activity cases. Moreover, the best-fit amplitudes grow generally larger as the temperature ranges get increasingly narrower. The only exception to this trend is found for the increasingly cooler temperature ranges, where the amplitude of the RVs extracted with the $\SI{4000}{}{-}\SI{4750}{K}$ spectral range reaches the minimum value (discussed further in Sect.~\ref{Sect:4.3}).

The time scale of the evolution of the periodic signal, $\tau$, is overall derived to be between 30 and 40 days, in accordance with previous literature values \citep[e.g.,][]{Camacho+2022}. We remark that $\tau$ is generally invariant to changes in both activity level and temperature range. This property can be particularly useful to, for example, model high- and low-activity seasons of the same star with two different quasi-periodic kernels at the same time. \cite{Klein+2024} have in fact shown that the covariance form of the solar activity in the RV time series does not stay constant, but changes throughout the solar cycle. Thus, increasing the analysed baseline to lengths of time comparable with the magnetic cycle does not improve the GP fit, and does not yield precise enough models for the hunt of small exoplanets. Possible solutions to this problem would be to either use a non-stationary version of the quasi-periodic kernel, or to model different sections of the solar data at the same time but with different quasi-periodic kernels. The latter technique would require $5\times N_{k}$ free parameters, in which $N_{k}$ is the number of kernels used for different time intervals. The analysis presented in this work has however shown that the time scale of the evolution of the periodic signal does not significantly change between the high- and low-activity seasons, which in the context of this multiple-kernel analysis would allow $\tau$ to stay constant in time, reducing the number of free parameters to $(4\times N_{k})+1$.

For the rotation period, $P_{\mathrm{rot}}$, the posteriors of the high-activity temperature ranges tend to be more clearly peaked than their low-activity counterparts, and a solar period can be more precisely derived. This is expected, as the larger variability at high activity is mostly generated by the presence of active regions and is thus more strongly modulated by the rotation of the star than the RV variations at low activity, which are instead dominated by other non-rotationally modulated effects \citep{Lakeland+2024}. Some of the extreme temperature ranges are the exception to this rule, showing clearly multi-peaked distributions. Although mostly within $1\sigma$ uncertainties, the derived $P_{\mathrm{rot}}$ for the high-activity cases are generally larger than the Carrington solar rotation period. On the other hand, the low-activity cases all prefer shorter periods, but have wider distributions and therefore larger uncertainties. This behaviour is expected, as during the descending phase of the solar cycle minimum, active regions tend to form at lower latitudes, which correspond to shorter rotational periods due to the Sun's differential rotation. Nevertheless, this trend cannot be considered statistically relevant due to the large overlap of ${\pm}1\sigma$ values. Significantly longer RV time series or more informative priors may be required to reproduce the detection of differential rotation proposed by \cite{Klein+2024}.

The distributions of the inverse harmonic complexity, $\mu$, tend to be biased toward lower values during high activity and and higher values during low activity. As a reminder, in this kernel formulation, larger $\mu$ values result in smoother GP functions in-between periods. $\mu$ is often related to the geometry or filling factors of active regions on the stellar surface \citep{Nicholson&Aigrain2022}. We however caution against assuming that larger inverse harmonic complexity values always describe clearer stellar surfaces with fewer or smaller active regions. \cite{Klein+2024} in fact note that a large co-moving spot or facula group at heightened activity might be simpler to model compared to several small and scattered magnetic regions at less active phases. Similar active region filling factors can also impact the best-fit value of $\mu$ differently based on their latitudinal distribution. In most cases, during low activity, the $\mu$ posteriors show long tails towards the upper boundary of the allowed parameter space. Inspecting this behaviour, by contextualising it with the other hyperparameters, shows that these larger $\mu$ values correspond to generally lower $P_{\mathrm{rot}}$ values. The GP is thus attempting to describe the complexity of the RV variability within the solar rotation by defining a lower $P_{\mathrm{rot}}$ with a smoother curve in-between periods.

Finally, the jitter term, $\beta$, appears to be bimodal in several cases. This bimodality is degenerate in posterior space with all other hyperparameters. By this we mean that $\beta$ values lying in both peaks of the jitter term posterior always belong to the same peak in the posteriors of all other GP hyperparameters. We attribute these double-peaks to MCMC samplers that are either able to include some of the left-over signal (see Sect.~\ref{Sect:4.3}) in the GP model, or omit it in favour of increased white noise. As a general rule for a time series well and fully described by a GP model with the selected covariance form, the value of the jitter term should be comparable to the inherent precision of the data. Too small $\beta$ values indicate overfitting (often via the absorption of white noise in the harmonic complexity); too large $\beta$ values suggest that the chosen kernel formulation is not able to fully describe the covariance of all the signals present in the analysed dataset. In the context of this work, the posteriors of the jitter are similar across both formation temperature range (with the exception of the hottest and coldest bins in which the uncertainties are significantly larger) and across activity level. These results point to a similar model precision floor across most considered time series.

\subsection{Best-fit models after GP regression}\label{Sect:4.3}

Figs.~\ref{Fig:03} and \ref{Fig:04} show the high- and low-activity RV time series, respectively, together with their \acrfull*{RMS} values, best-fit GP model, and their residuals, following the same colour convention as in Fig.~\ref{Fig:01}. The figures also show the \acrlong*{GLS} \citep[GLS;][]{Lomb1976,Scargle1982,Zechmeister&Kurster2009} periodograms of the observations, the models, and the residuals. The GP models used for the detrending utilise the $50^{\mathrm{th}}$ percentile values of the hyperparameter distributions in Fig.~\ref{Fig:02}.

For the high-activity time interval, we find that the $\SI{4000}{}{-}\SI{4750}{K}$ range produces the smallest RV RMS for the uncorrected RV curves; in other words, minimum RV RMS is not attained at maximum RV content, i.e. when the entire spectrum is utilised to compute the RV. We also remark a clear transition for the strongest peak in the periodogram of the observed RVs, from the full solar rotation period at hotter temperatures, to half the solar rotation period at cooler temperatures. This transition is likewise found by \cite{AlMoulla+2022}, who argued that this could be due to a change in balance between the convective and photometric RV components (see Sect.~\ref{Sect:4.4}), whose signatures favour different harmonics of the stellar rotation period. The uncorrected RV curves all have RMS values exceeding $\SI{1}{\meter\per\second}$, however, after subtracting their respective best-fit GP models, their dispersions are all brought to the sub-$\SI{}{\meter\per\second}$ level. The RV RMS of the $\SI{4000}{}{-}\SI{4750}{K}$ range remain the smallest both for the observed RVs and the corrected residuals. These results indicate that the RVs derived from this particular line formation temperature range are, if only by a small margin, the least affected by solar variability, and that the covariance of their activity signal is particularly well-described by a GP with a QP kernel. This line formation temperature range can therefore be a promising avenue for RV extraction of the Sun, and perhaps Sun-like stars, yielding RV time series with intrinsically lower scatter more suitable for, e.g., recovery of planetary signals, granulation modelling, or further GP analysis. We also remark that while the two coolest temperature ranges present the largest RMS reduction after model subtraction, their residual RMS is still larger than that of the $\SI{4000}{}{-}\SI{4750}{K}$ range.

For the low-activity time interval, we find generally similar RV RMS results as for the high-activity interval, with the exception that the uncorrected RV data start off with lower RMS values. Once again, the RMS values of the RVs from the two coldest temperature ranges are reduced the most via GP model subtraction, but the lowest RMS value in the residuals is obtained with the data extracted from the third coolest ($\SI{4000}{}{-}\SI{4750}{K}$) temperature range, mimicking the behaviour of the high-activity datasets. The solar variability present in the RVs from this line formation temperature range has therefore been shown to be the better described by GPs with QP kernels over both high and low activity levels. The RVs derived from this range can thus be used to isolate some of the solar characteristics driving the quasi-periodic variability more reliably than by using the RVs extracted from the entire spectrum.

In all cases, the residual RVs computed as the subtraction between the observed temperature-dependent RVs and the best-fit GP model all generate overall flat periodograms. In particular, the GPs successfully model out the long-period signals in the data for both activity levels and all temperature bins. With the exception of some high-frequency signals, the only relevant peak left in most of the periodograms of the residuals is shared between most time series at $\sim$8 to 10 days. This common peak can also be partially found in the periodogram of the observed RVs. Given its consistency between datasets, it is unlikely to be a product of improper GP fitting. This points to a possible uncorrected signal (at least partially periodic in nature) present in the data. Its periodicity is comparable to the third harmonic of the solar rotation period, or to the timescale of r-modes as described by \cite{Lanza+2019}. The determination of the exact source of this signal is however beyond the scope of this work.

We also note that for both the high- and low-activity intervals, the GP models are unable to lower the residual RV RMS values below ${\sim}\SI{50}{\centi\meter\per\second}$. While part of these residual RMS can be attributed to the inherent white noise present in the data, this common RMS floor value also points to uncorrected variability similarly present in most temperature ranges and for both activity levels. This is some solar activity component not currently accounted for in the GP model. The covariance form chosen for this analysis, the QP kernel, has been shown to successfully model rotationally modulated signals (usually mainly attributed to the presence of active regions rotating in and out of sight of the observer). However, all RV variability not inherently periodic or modulated by a different periodicity than the stellar rotation is not included in the model. The RMS values of most residual time series are comparable to the expected dispersion caused by convective motions in the solar surface granules, referred to as granulation and supergranulation depending on the size and timescale of the granule structures; their velocimetric signal have been measured for Sun-like stars \citep{Dumusque+2011}, the Sun itself \citep{AlMoulla+2023,Lakeland+2024}, and simulated for Sun-like conditions \citep{Meunier+2015,Palumbo+2024}. This result highlights the need of further analysis of these leftover signals and of more complex kernel formulations to fully describe the stellar activity.

\begin{figure}
	\includegraphics[width=\linewidth]{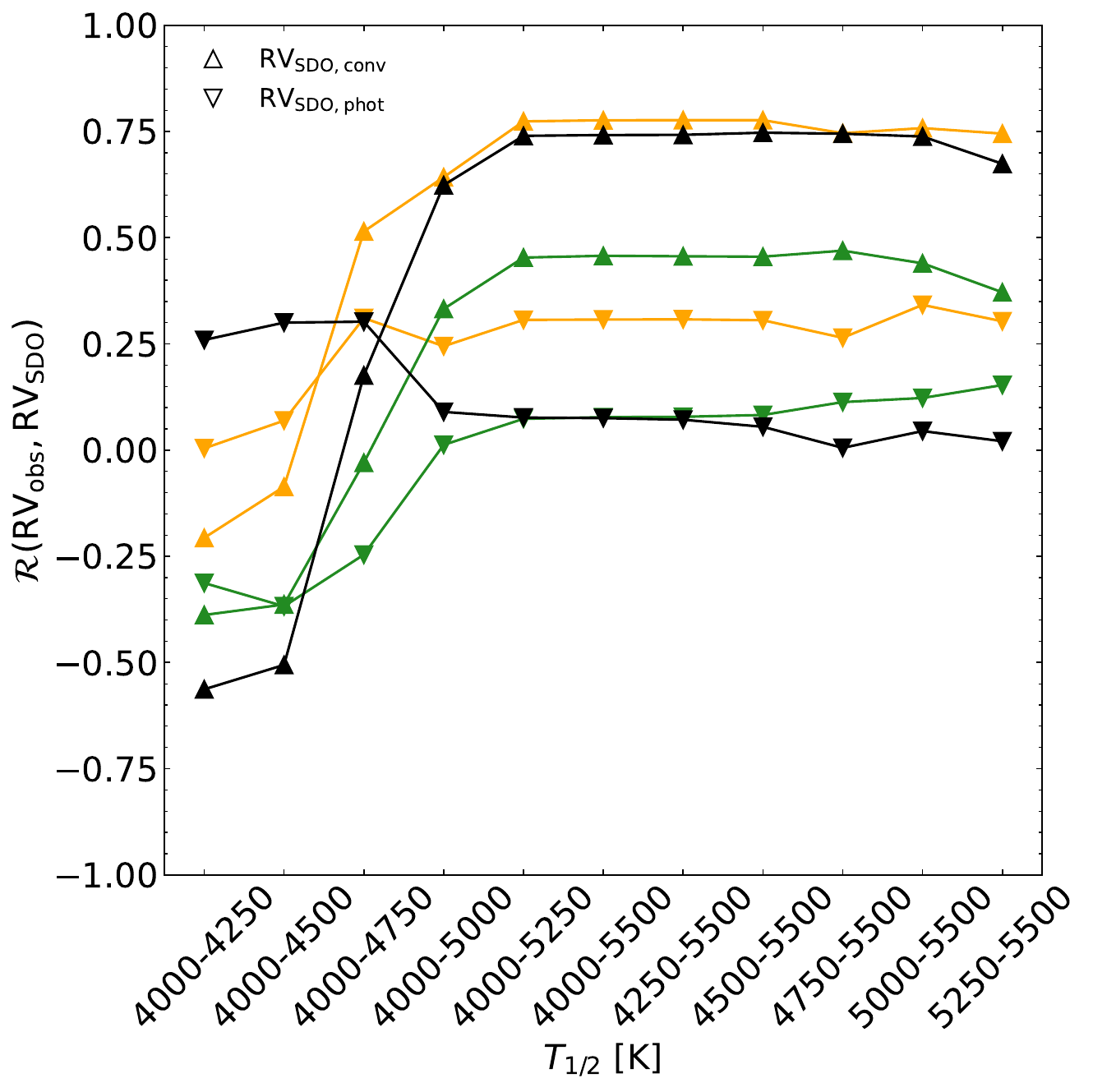}
    \caption{Pearson correlations coefficients between the observed HARPS-N RVs, $\mathrm{RV}_{\mathrm{obs}}$, and RVs extracted from SDO Dopplergrams, $\mathrm{RV}_{\mathrm{SDO}}$. The upward and downward triangles show the correlations for the RV variations from the inhibition of convective blueshift, $\mathrm{RV}_{\mathrm{SDO,conv}}$, and from the photometric contrast, $\mathrm{RV}_{\mathrm{SDO,phot}}$, by magnetically active surface regions, respectively. The black, orange, and green curves correspond to the total time series, the high activity time interval, and the low activity time interval, respectively.}
    \label{Fig:05}
\end{figure}

\subsection{Comparison with SDO Dopplergrams}\label{Sect:4.4}

Disk-resolved images of the Sun taken by the \acrlong*{HMI} \citep[HMI;][]{Scherrer+2012,Schou+2012} onboard the \acrlong*{SDO} \citep[SDO;][]{Pesnell+2012} provide a valuable complement to the disk-integrated spectra from which we derive our RVs. SDO Dopplergrams allow us to evaluate the various components that add up to the total disk-integrated RVs. For example, one can determine the RV contribution from specific types of magnetically active regions, i.e., dark spots or bright faculae, or from specific effects by which these active regions interact with the surface flows, i.e., either their inhibition of the underlying convective blueshift (hereafter referred to as the convective component) or their perturbation to the photometric balance between the red- and blue-shifted hemispheres due to the Sun's rotation (hereafter the photometric component).

We make use of the \texttt{SolAster}\footnote{ \url{https://github.com/tamarervin/SolAster}} \citep{Ervin+2022} code, which follows the methodology of \cite{Haywood+2016}, \cite{Milbourne+2019} and \cite{Haywood+2022} to extract the convective and photometric RV components of the disk-integrated Dopplergrams. These velocities are obtained from SDO frames with an exposure time of $\SI{720}{\second}$ taken at UTC 12:00:00 every day of the observed time span with HARPS-N. The RV components are thereafter linearly interpolated onto the exact same time stamps as the observations.

In order to investigate whether the RV signal at certain formation temperature ranges is primarily explained by specific effects, we compute the Pearson correlation coefficients between the temperature-dependent RV time series and the contemporaneous convective and photometric RVs from SDO. The results are presented in Fig.~\ref{Fig:05}, which shows the evolution of the correlation with temperature for the total time span as well as the high- and low-activity time intervals from Fig.~\ref{Fig:01}. We find that the observed RVs from spectral segments formed at higher temperatures strongly correlate with the convective RV contribution when considering the total data set and the high-activity time interval. This correlation decreases at lower formation temperatures and transitions to a relatively strong anti-correlation at the coolest temperature range. During low activity, the correlations are generally weaker, albeit showing a similar trend with formation temperature. For the photometric RV contribution, we do not find comparably strong correlations within any temperature range. Although the correlation increases from near-zero at the hottest temperature range to higher values at the cooler ranges, it stagnates at a relatively low ${\sim}0.3$ value.

These correlations strengthen the consensus that the inhibition of convective blueshift (the convective component) is the dominant RV component in Sun-like variability \citep{Meunier+2010}, and that the dominant variations occur in spectral segments formed at the hotter end of the line-forming temperature range \citep{AlMoulla+2022}. The RV variations from spectral segments formed at the coolest temperature ranges were proposed by \cite{AlMoulla+2022} to be due to the inhibition of convective redshift, which in turn could be caused by an intensity reversal of the granules and intergranular lanes \citep{Janssen&Cauzzi2006}. This claim was brought forward because the RVs related to cooler temperature ranges were found to be anti-correlated with the $S$ index; we find further evidence by the fact that these RV variations are anti-correlated with the convective RV component specifically. However, the somewhat weaker absolute correlation compared to the opposite temperature extreme, as well as the coincidental increase in correlation with the photometric RV component, indicates that a reversed granulation signal might not be a full description of the physical processes governing the cooler layers of the photosphere.

\section{Conclusions}\label{Sect:5}

In this paper we model RVs measured from spectral segments formed at various photospheric temperatures with quasi-periodic GPs in order to better understand how stellar activity signals vary with formation temperature, and whether the model parameters could inform us about the origin or characteristics of the varying RV signatures. We make use of solar disk-integrated spectra which have been corrected for the influence of Solar System planets and instrumental systematics, predominantly isolating the imprint of stellar variability.

We perform GP regression with MCMC likelihood optimisation to 11 formation temperature ranges for two time series corresponding to a high- and low-activity time interval. Broadly claimed for the kernel hyperparameters, we find that the solar rotation period is better constrained at heightened activity levels, that the evolution timescale is invariant to both activity level and temperature changes, that the inverse harmonic complexity is biased to smaller values at high activity, that the amplitude grows with narrower temperature range, and that the jitter is bimodal due to its occasional absorption of a potentially coherent (super-)granulation signal.

Interestingly, we find that the formation temperature range which leads to the smallest RV dispersion is not the one corresponding to the entire spectral range ($\SI{4000}{}{-}\SI{5500}{K}$) but rather an intermediately cool range ($\SI{4000}{}{-}\SI{4750}{K}$). This range has the smallest RV RMS for both its uncorrected RVs and its RV residuals after subtracting the GP model.

Importantly, we also find that for both activity level intervals and for all temperature ranges, the GP models are unable to lower the RMS of the residual RVs below ${\sim}\SI{50}{\centi\meter\per\second}$. This shared floor highlights the presence of currently unmodelled stellar signals present in the data comparable with those generated by convecting motions, such as supergranulation \citep[e.g.,][]{Meunier+2015}.

When comparing our temperature-dependent RVs with contemporaneous RV components extracted from SDO Dopplergrams, we find that the RV component due to the inhibition of convective blueshift by large magnetically active regions strongly correlates with observed RVs measured at spectral ranges formed at hotter temperatures. This correlation decreases toward cooler temperature ranges, becoming an anti-correlation at the opposite extreme possibly due to an intensity reversal in the upper photosphere leading to the likewise reversal of the observed flows. We hope that future investigations could make use of magnetohydrodynamical simulations of solar convection, in order to compare whether any drastic changes in the contrast between convective granules and intergranular lanes occur at the same photospheric temperature ranges as inferred from our method.

We believe our results could address several important aspects within the research field of stellar activity: understanding the behaviours of model parameters at different activity phases could guide and motivate their usage for lesser-monitored stars than the Sun; finding spectral segments which are inherently less susceptible to effects of stellar variability or more easily modelled could alleviate the need for extensive detrending techniques; and connecting observed RVs to those independently obtained from disk-resolved Dopplergrams could bring the field closer to a complete picture of the stellar activity processes which govern the surface flows of Sun-like stars.

\section*{Acknowledgements}

We are deeply thankful to the anonymous reviewers whose suggestions and comments have helped improve the quality of our analysis and the clarity of our manuscript. This work was supported by the Science and Technology Funding Council ST/Y002334/1. This work has been carried out within the framework of the National Centre of Competence in Research PlanetS supported by the Swiss National Science Foundation under grants 51NF40\_182901 and 51NF40\_205606. This project has received funding from the European Research Council (ERC) under the European Union’s Horizon 2020 research and innovation program (grant agreement SCORE No. 851555). This work has made use of the VALD database, operated at Uppsala University, the Institute of Astronomy RAS in Moscow, and the University of Vienna. This project was developed in part at the 2023 Sun-as-a-Star Workshop, hosted by the Center for Computational Astrophysics of the Flatiron Institute in New York City. The authors thank Baptiste Klein, Ben Lakeland, Eric Ford, Megan Bedell, Michael Cretignier, Michael Palumbo, Niamh O'Sullivan, Rapha{\"e}lle Haywood, Ryan Rubenzahl, and Xavier Dumusque for fruitful discussions.

\section*{Data Availability}

The first three years of HARPS-N solar spectra are publicly available at the \acrfull*{DACE}:\\\url{https://dace.unige.ch/sun}\\The SDO Dopplergrams are publicly available at the \acrfull*{JSOC}:\\\url{http://jsoc.stanford.edu}



\balance
\bibliographystyle{mnras}
\bibliography{References} 

\begin{thebibliography}{}
\makeatletter
\relax
\def\mn@urlcharsother{\let\do\@makeother \do\$\do\&\do\#\do\^\do\_\do\%\do\~}
\def\mn@doi{\begingroup\mn@urlcharsother \@ifnextchar [ {\mn@doi@} {\mn@doi@[]}}
\def\mn@doi@[#1]#2{\def\@tempa{#1}\ifx\@tempa\@empty \href {http://dx.doi.org/#2} {doi:#2}\else \href {http://dx.doi.org/#2} {#1}\fi \endgroup}
\def\mn@eprint#1#2{\mn@eprint@#1:#2::\@nil}
\def\mn@eprint@arXiv#1{\href {http://arxiv.org/abs/#1} {{\tt arXiv:#1}}}
\def\mn@eprint@dblp#1{\href {http://dblp.uni-trier.de/rec/bibtex/#1.xml} {dblp:#1}}
\def\mn@eprint@#1:#2:#3:#4\@nil{\def\@tempa {#1}\def\@tempb {#2}\def\@tempc {#3}\ifx \@tempc \@empty \let \@tempc \@tempb \let \@tempb \@tempa \fi \ifx \@tempb \@empty \def\@tempb {arXiv}\fi \@ifundefined {mn@eprint@\@tempb}{\@tempb:\@tempc}{\expandafter \expandafter \csname mn@eprint@\@tempb\endcsname \expandafter{\@tempc}}}

\bibitem[\protect\citeauthoryear{{Al Moulla}, {Dumusque}, {Cretignier}, {Zhao}  \& {Valenti}}{{Al Moulla} et~al.}{2022}]{AlMoulla+2022}
{Al Moulla} K.,  {Dumusque} X.,  {Cretignier} M.,  {Zhao} Y.,   {Valenti} J.~A.,  2022, \mn@doi [\aap] {10.1051/0004-6361/202243276}, \href {https://ui.adsabs.harvard.edu/abs/2022A&A...664A..34A} {664, A34}

\bibitem[\protect\citeauthoryear{{Al Moulla}, {Dumusque}, {Figueira}, {Lo Curto}, {Santos}  \& {Wildi}}{{Al Moulla} et~al.}{2023}]{AlMoulla+2023}
{Al Moulla} K.,  {Dumusque} X.,  {Figueira} P.,  {Lo Curto} G.,  {Santos} N.~C.,   {Wildi} F.,  2023, \mn@doi [\aap] {10.1051/0004-6361/202244663}, \href {https://ui.adsabs.harvard.edu/abs/2023A&A...669A..39A} {669, A39}

\bibitem[\protect\citeauthoryear{{Al Moulla}, {Dumusque}  \& {Cretignier}}{{Al Moulla} et~al.}{2024}]{AlMoulla+2024}
{Al Moulla} K.,  {Dumusque} X.,   {Cretignier} M.,  2024, \mn@doi [\aap] {10.1051/0004-6361/202348150}, \href {https://ui.adsabs.harvard.edu/abs/2024A&A...683A.106A} {683, A106}

\bibitem[\protect\citeauthoryear{{Anglada-Escud{\'e}} \& {Butler}}{{Anglada-Escud{\'e}} \& {Butler}}{2012}]{Anglada-Escude&Butler2012}
{Anglada-Escud{\'e}} G.,  {Butler} R.~P.,  2012, \mn@doi [\apjs] {10.1088/0067-0049/200/2/15}, \href {https://ui.adsabs.harvard.edu/abs/2012ApJS..200...15A} {200, 15}

\bibitem[\protect\citeauthoryear{{Artigau} et~al.,}{{Artigau} et~al.}{2022}]{Artigau+2022}
{Artigau} {\'E}.,  et~al., 2022, \mn@doi [\aj] {10.3847/1538-3881/ac7ce6}, \href {https://ui.adsabs.harvard.edu/abs/2022AJ....164...84A} {164, 84}

\bibitem[\protect\citeauthoryear{{Asplund}, {Grevesse}, {Sauval}  \& {Scott}}{{Asplund} et~al.}{2009}]{Asplund+2009}
{Asplund} M.,  {Grevesse} N.,  {Sauval} A.~J.,   {Scott} P.,  2009, \mn@doi [\araa] {10.1146/annurev.astro.46.060407.145222}, \href {https://ui.adsabs.harvard.edu/abs/2009ARA&A..47..481A} {47, 481}

\bibitem[\protect\citeauthoryear{{Baranne} et~al.,}{{Baranne} et~al.}{1996}]{Baranne+1996}
{Baranne} A.,  et~al., 1996, \aaps, \href {https://ui.adsabs.harvard.edu/abs/1996A&AS..119..373B} {119, 373}

\bibitem[\protect\citeauthoryear{{Barros}, {Demangeon}, {D{\'\i}az}, {Cabrera}, {Santos}, {Faria}  \& {Pereira}}{{Barros} et~al.}{2020}]{Barros+2020}
{Barros} S.~C.~C.,  {Demangeon} O.,  {D{\'\i}az} R.~F.,  {Cabrera} J.,  {Santos} N.~C.,  {Faria} J.~P.,   {Pereira} F.,  2020, \mn@doi [\aap] {10.1051/0004-6361/201936086}, \href {https://ui.adsabs.harvard.edu/abs/2020A&A...634A..75B} {634, A75}

\bibitem[\protect\citeauthoryear{{Camacho}, {Faria}  \& {Viana}}{{Camacho} et~al.}{2022}]{Camacho+2022}
{Camacho} J.~D.,  {Faria} J.~P.,   {Viana} P.~T.~P.,  2022, arXiv e-prints, \href {https://ui.adsabs.harvard.edu/abs/2022arXiv220506627C} {p. arXiv:2205.06627}

\bibitem[\protect\citeauthoryear{{Collier Cameron} et~al.,}{{Collier Cameron} et~al.}{2019}]{CollierCameron+2019}
{Collier Cameron} A.,  et~al., 2019, \mn@doi [\mnras] {10.1093/mnras/stz1215}, \href {https://ui.adsabs.harvard.edu/abs/2019MNRAS.487.1082C} {487, 1082}

\bibitem[\protect\citeauthoryear{{Cosentino} et~al.,}{{Cosentino} et~al.}{2012}]{Cosentino+2012}
{Cosentino} R.,  et~al., 2012, in {McLean} I.~S.,  {Ramsay} S.~K.,   {Takami} H.,  eds,  Society of Photo-Optical Instrumentation Engineers (SPIE) Conference Series Vol. 8446, Ground-based and Airborne Instrumentation for Astronomy IV. p. 84461V, \mn@doi{10.1117/12.925738}

\bibitem[\protect\citeauthoryear{{Cosentino} et~al.,}{{Cosentino} et~al.}{2014}]{Cosentino+2014}
{Cosentino} R.,  et~al., 2014, in {Ramsay} S.~K.,  {McLean} I.~S.,   {Takami} H.,  eds,  Society of Photo-Optical Instrumentation Engineers (SPIE) Conference Series Vol. 9147, Ground-based and Airborne Instrumentation for Astronomy V. p. 91478C, \mn@doi{10.1117/12.2055813}

\bibitem[\protect\citeauthoryear{{Costes} et~al.,}{{Costes} et~al.}{2021}]{Costes+2021}
{Costes} J.~C.,  et~al., 2021, \mn@doi [\mnras] {10.1093/mnras/stab1183}, \href {https://ui.adsabs.harvard.edu/abs/2021MNRAS.505..830C} {505, 830}

\bibitem[\protect\citeauthoryear{{Crass} et~al.,}{{Crass} et~al.}{2021}]{Crass+2021}
{Crass} J.,  et~al., 2021, \mn@doi [arXiv e-prints] {10.48550/arXiv.2107.14291}, \href {https://ui.adsabs.harvard.edu/abs/2021arXiv210714291C} {p. arXiv:2107.14291}

\bibitem[\protect\citeauthoryear{{Cretignier}, {Dumusque}, {Hara}  \& {Pepe}}{{Cretignier} et~al.}{2021}]{Cretignier+2021}
{Cretignier} M.,  {Dumusque} X.,  {Hara} N.~C.,   {Pepe} F.,  2021, \mn@doi [\aap] {10.1051/0004-6361/202140986}, \href {https://ui.adsabs.harvard.edu/abs/2021A&A...653A..43C} {653, A43}

\bibitem[\protect\citeauthoryear{{Cretignier}, {Dumusque}, {Aigrain}  \& {Pepe}}{{Cretignier} et~al.}{2023}]{Cretignier+2023}
{Cretignier} M.,  {Dumusque} X.,  {Aigrain} S.,   {Pepe} F.,  2023, \mn@doi [\aap] {10.1051/0004-6361/202347232}, \href {https://ui.adsabs.harvard.edu/abs/2023A&A...678A...2C} {678, A2}

\bibitem[\protect\citeauthoryear{{Cretignier}, {Pietrow}  \& {Aigrain}}{{Cretignier} et~al.}{2024}]{Cretignier+2024}
{Cretignier} M.,  {Pietrow} A.~G.~M.,   {Aigrain} S.,  2024, \mn@doi [\mnras] {10.1093/mnras/stad3292}, \href {https://ui.adsabs.harvard.edu/abs/2024MNRAS.527.2940C} {527, 2940}

\bibitem[\protect\citeauthoryear{{Dalal} et~al.,}{{Dalal} et~al.}{2024}]{Dalal+2024}
{Dalal} S.,  et~al., 2024, \mn@doi [\mnras] {10.1093/mnras/stae1367}, \href {https://ui.adsabs.harvard.edu/abs/2024MNRAS.531.4464D} {531, 4464}

\bibitem[\protect\citeauthoryear{{Desort}, {Lagrange}, {Galland}, {Udry}  \& {Mayor}}{{Desort} et~al.}{2007}]{Desort+2007}
{Desort} M.,  {Lagrange} A.~M.,  {Galland} F.,  {Udry} S.,   {Mayor} M.,  2007, \mn@doi [\aap] {10.1051/0004-6361:20078144}, \href {https://ui.adsabs.harvard.edu/abs/2007A&A...473..983D} {473, 983}

\bibitem[\protect\citeauthoryear{{Dumusque}}{{Dumusque}}{2018}]{Dumusque2018}
{Dumusque} X.,  2018, \mn@doi [\aap] {10.1051/0004-6361/201833795}, \href {https://ui.adsabs.harvard.edu/abs/2018A&A...620A..47D} {620, A47}

\bibitem[\protect\citeauthoryear{{Dumusque}, {Udry}, {Lovis}, {Santos}  \& {Monteiro}}{{Dumusque} et~al.}{2011}]{Dumusque+2011}
{Dumusque} X.,  {Udry} S.,  {Lovis} C.,  {Santos} N.~C.,   {Monteiro} M.~J.~P.~F.~G.,  2011, \mn@doi [\aap] {10.1051/0004-6361/201014097}, \href {https://ui.adsabs.harvard.edu/abs/2011A&A...525A.140D} {525, A140}

\bibitem[\protect\citeauthoryear{{Dumusque} et~al.,}{{Dumusque} et~al.}{2015}]{Dumusque+2015}
{Dumusque} X.,  et~al., 2015, \mn@doi [\apjl] {10.1088/2041-8205/814/2/L21}, \href {https://ui.adsabs.harvard.edu/abs/2015ApJ...814L..21D} {814, L21}

\bibitem[\protect\citeauthoryear{{Dumusque} et~al.,}{{Dumusque} et~al.}{2021}]{Dumusque+2021}
{Dumusque} X.,  et~al., 2021, \mn@doi [\aap] {10.1051/0004-6361/202039350}, \href {https://ui.adsabs.harvard.edu/abs/2021A&A...648A.103D} {648, A103}

\bibitem[\protect\citeauthoryear{{Ervin} et~al.,}{{Ervin} et~al.}{2022}]{Ervin+2022}
{Ervin} T.,  et~al., 2022, \mn@doi [\aj] {10.3847/1538-3881/ac67e6}, \href {https://ui.adsabs.harvard.edu/abs/2022AJ....163..272E} {163, 272}

\bibitem[\protect\citeauthoryear{{Fischer} et~al.,}{{Fischer} et~al.}{2016}]{Fischer+2016}
{Fischer} D.~A.,  et~al., 2016, \mn@doi [PASP] {10.1088/1538-3873/128/964/066001}, \href {https://ui.adsabs.harvard.edu/abs/2016PASP..128f6001F} {128, 066001}

\bibitem[\protect\citeauthoryear{{Foreman-Mackey}, {Agol}, {Ambikasaran}  \& {Angus}}{{Foreman-Mackey} et~al.}{2017}]{Foreman-Mackey+2017}
{Foreman-Mackey} D.,  {Agol} E.,  {Ambikasaran} S.,   {Angus} R.,  2017, \mn@doi [\apj] {10.3847/1538-3881/aa9332}, \href {https://ui.adsabs.harvard.edu/abs/2017AJ....154..220F} {154, 220}

\bibitem[\protect\citeauthoryear{{Gray}}{{Gray}}{2008}]{Gray2008}
{Gray} D.~F.,  2008, {The Observation and Analysis of Stellar Photospheres}, 3 edn.
Cambridge University Press

\bibitem[\protect\citeauthoryear{{Gustafsson}, {Edvardsson}, {Eriksson}, {J{\o}rgensen}, {Nordlund}  \& {Plez}}{{Gustafsson} et~al.}{2008}]{Gustafsson+2008}
{Gustafsson} B.,  {Edvardsson} B.,  {Eriksson} K.,  {J{\o}rgensen} U.~G.,  {Nordlund} {\r{A}}.,   {Plez} B.,  2008, \mn@doi [\aap] {10.1051/0004-6361:200809724}, \href {https://ui.adsabs.harvard.edu/abs/2008A&A...486..951G} {486, 951}

\bibitem[\protect\citeauthoryear{{Haywood} et~al.,}{{Haywood} et~al.}{2014}]{Haywood+2014}
{Haywood} R.~D.,  et~al., 2014, \mn@doi [\mnras] {10.1093/mnras/stu1320}, \href {https://ui.adsabs.harvard.edu/abs/2014MNRAS.443.2517H} {443, 2517}

\bibitem[\protect\citeauthoryear{{Haywood} et~al.,}{{Haywood} et~al.}{2016}]{Haywood+2016}
{Haywood} R.~D.,  et~al., 2016, \mn@doi [\mnras] {10.1093/mnras/stw187}, \href {https://ui.adsabs.harvard.edu/abs/2016MNRAS.457.3637H} {457, 3637}

\bibitem[\protect\citeauthoryear{{Haywood} et~al.,}{{Haywood} et~al.}{2022}]{Haywood+2022}
{Haywood} R.~D.,  et~al., 2022, \mn@doi [\apj] {10.3847/1538-4357/ac7c12}, \href {https://ui.adsabs.harvard.edu/abs/2022ApJ...935....6H} {935, 6}

\bibitem[\protect\citeauthoryear{{Janssen} \& {Cauzzi}}{{Janssen} \& {Cauzzi}}{2006}]{Janssen&Cauzzi2006}
{Janssen} K.,  {Cauzzi} G.,  2006, \mn@doi [\aap] {10.1051/0004-6361:20054310}, \href {https://ui.adsabs.harvard.edu/abs/2006A&A...450..365J} {450, 365}

\bibitem[\protect\citeauthoryear{{John}, {Collier Cameron}  \& {Wilson}}{{John} et~al.}{2022}]{John+2022}
{John} A.~A.,  {Collier Cameron} A.,   {Wilson} T.~G.,  2022, \mn@doi [\mnras] {10.1093/mnras/stac1814}, \href {https://ui.adsabs.harvard.edu/abs/2022MNRAS.515.3975J} {515, 3975}

\bibitem[\protect\citeauthoryear{{Klein} et~al.,}{{Klein} et~al.}{2024}]{Klein+2024}
{Klein} B.,  et~al., 2024, \mn@doi [\mnras] {10.1093/mnras/stae1313}, \href {https://ui.adsabs.harvard.edu/abs/2024MNRAS.531.4238K} {531, 4238}

\bibitem[\protect\citeauthoryear{{Kupka}, {Ryabchikova}, {Piskunov}, {Stempels}  \& {Weiss}}{{Kupka} et~al.}{2000}]{Kupka+2000}
{Kupka} F.~G.,  {Ryabchikova} T.~A.,  {Piskunov} N.~E.,  {Stempels} H.~C.,   {Weiss} W.~W.,  2000, \mn@doi [Baltic Astronomy] {10.1515/astro-2000-0420}, \href {https://ui.adsabs.harvard.edu/abs/2000BaltA...9..590K} {9, 590}

\bibitem[\protect\citeauthoryear{{Lakeland} et~al.,}{{Lakeland} et~al.}{2024}]{Lakeland+2024}
{Lakeland} B.~S.,  et~al., 2024, \mn@doi [\mnras] {10.1093/mnras/stad3723}, \href {https://ui.adsabs.harvard.edu/abs/2024MNRAS.527.7681L} {527, 7681}

\bibitem[\protect\citeauthoryear{{Lanza}, {Gizon}, {Zaqarashvili}, {Liang}  \& {Rodenbeck}}{{Lanza} et~al.}{2019}]{Lanza+2019}
{Lanza} A.~F.,  {Gizon} L.,  {Zaqarashvili} T.~V.,  {Liang} Z.~C.,   {Rodenbeck} K.,  2019, \mn@doi [\aap] {10.1051/0004-6361/201834712}, \href {https://ui.adsabs.harvard.edu/abs/2019A&A...623A..50L} {623, A50}

\bibitem[\protect\citeauthoryear{{Lomb}}{{Lomb}}{1976}]{Lomb1976}
{Lomb} N.~R.,  1976, \mn@doi [\apss] {10.1007/BF00648343}, \href {https://ui.adsabs.harvard.edu/abs/1976Ap&SS..39..447L} {39, 447}

\bibitem[\protect\citeauthoryear{{Meunier}}{{Meunier}}{2021}]{Meunier2021}
{Meunier} N.,  2021, \mn@doi [arXiv e-prints] {10.48550/arXiv.2104.06072}, \href {https://ui.adsabs.harvard.edu/abs/2021arXiv210406072M} {p. arXiv:2104.06072}

\bibitem[\protect\citeauthoryear{{Meunier}, {Desort}  \& {Lagrange}}{{Meunier} et~al.}{2010}]{Meunier+2010}
{Meunier} N.,  {Desort} M.,   {Lagrange} A.~M.,  2010, \mn@doi [\aap] {10.1051/0004-6361/200913551}, \href {https://ui.adsabs.harvard.edu/abs/2010A&A...512A..39M} {512, A39}

\bibitem[\protect\citeauthoryear{{Meunier}, {Lagrange}, {Borgniet}  \& {Rieutord}}{{Meunier} et~al.}{2015}]{Meunier+2015}
{Meunier} N.,  {Lagrange} A.~M.,  {Borgniet} S.,   {Rieutord} M.,  2015, \mn@doi [\aap] {10.1051/0004-6361/201525721}, \href {https://ui.adsabs.harvard.edu/abs/2015A&A...583A.118M} {583, A118}

\bibitem[\protect\citeauthoryear{{Milbourne} et~al.,}{{Milbourne} et~al.}{2019}]{Milbourne+2019}
{Milbourne} T.~W.,  et~al., 2019, \mn@doi [\apj] {10.3847/1538-4357/ab064a}, \href {https://ui.adsabs.harvard.edu/abs/2019ApJ...874..107M} {874, 107}

\bibitem[\protect\citeauthoryear{{Nava} et~al.,}{{Nava} et~al.}{2022}]{Nava+2022}
{Nava} C.,  et~al., 2022, \mn@doi [\aj] {10.3847/1538-3881/ac3141}, \href {https://ui.adsabs.harvard.edu/abs/2022AJ....163...41N} {163, 41}

\bibitem[\protect\citeauthoryear{{Nicholson} \& {Aigrain}}{{Nicholson} \& {Aigrain}}{2022}]{Nicholson&Aigrain2022}
{Nicholson} B.~A.,  {Aigrain} S.,  2022, \mn@doi [\mnras] {10.1093/mnras/stac2097}, \href {https://ui.adsabs.harvard.edu/abs/2022MNRAS.515.5251N} {515, 5251}

\bibitem[\protect\citeauthoryear{{Palumbo}, {Ford}, {Gonzalez}, {Wright}, {Al Moulla}  \& {Schlichenmaier}}{{Palumbo} et~al.}{2024}]{Palumbo+2024}
{Palumbo} M.~L.,  {Ford} E.~B.,  {Gonzalez} E.~B.,  {Wright} J.~T.,  {Al Moulla} K.,   {Schlichenmaier} R.,  2024, \mn@doi [\aj] {10.3847/1538-3881/ad4c6d}, \href {https://ui.adsabs.harvard.edu/abs/2024AJ....168...46P} {168, 46}

\bibitem[\protect\citeauthoryear{{Pepe}, {Mayor}, {Galland}, {Naef}, {Queloz}, {Santos}, {Udry}  \& {Burnet}}{{Pepe} et~al.}{2002}]{Pepe+2002}
{Pepe} F.,  {Mayor} M.,  {Galland} F.,  {Naef} D.,  {Queloz} D.,  {Santos} N.~C.,  {Udry} S.,   {Burnet} M.,  2002, \mn@doi [\aap] {10.1051/0004-6361:20020433}, \href {https://ui.adsabs.harvard.edu/abs/2002A&A...388..632P} {388, 632}

\bibitem[\protect\citeauthoryear{{Pesnell}, {Thompson}  \& {Chamberlin}}{{Pesnell} et~al.}{2012}]{Pesnell+2012}
{Pesnell} W.~D.,  {Thompson} B.~J.,   {Chamberlin} P.~C.,  2012, \mn@doi [SolPhys] {10.1007/s11207-011-9841-3}, \href {https://ui.adsabs.harvard.edu/abs/2012SoPh..275....3P} {275, 3}

\bibitem[\protect\citeauthoryear{{Phillips} et~al.,}{{Phillips} et~al.}{2016}]{Phillips+2016}
{Phillips} D.~F.,  et~al., 2016, in {Navarro} R.,  {Burge} J.~H.,  eds,  Society of Photo-Optical Instrumentation Engineers (SPIE) Conference Series Vol. 9912, Advances in Optical and Mechanical Technologies for Telescopes and Instrumentation II. p. 99126Z, \mn@doi{10.1117/12.2232452}

\bibitem[\protect\citeauthoryear{{Piskunov} \& {Valenti}}{{Piskunov} \& {Valenti}}{2017}]{Piskunov&Valenti2017}
{Piskunov} N.,  {Valenti} J.~A.,  2017, \mn@doi [\aap] {10.1051/0004-6361/201629124}, \href {https://ui.adsabs.harvard.edu/abs/2017A&A...597A..16P} {597, A16}

\bibitem[\protect\citeauthoryear{{Piskunov}, {Kupka}, {Ryabchikova}, {Weiss}  \& {Jeffery}}{{Piskunov} et~al.}{1995}]{Piskunov+1995}
{Piskunov} N.~E.,  {Kupka} F.,  {Ryabchikova} T.~A.,  {Weiss} W.~W.,   {Jeffery} C.~S.,  1995, \aaps, \href {https://ui.adsabs.harvard.edu/abs/1995A&AS..112..525P} {112, 525}

\bibitem[\protect\citeauthoryear{{Queloz} et~al.,}{{Queloz} et~al.}{2001}]{Queloz+2001}
{Queloz} D.,  et~al., 2001, \mn@doi [\aap] {10.1051/0004-6361:20011308}, \href {https://ui.adsabs.harvard.edu/abs/2001A&A...379..279Q} {379, 279}

\bibitem[\protect\citeauthoryear{{Queloz} et~al.,}{{Queloz} et~al.}{2009}]{Queloz+2009}
{Queloz} D.,  et~al., 2009, \mn@doi [\aap] {10.1051/0004-6361/200913096}, \href {https://ui.adsabs.harvard.edu/abs/2009A&A...506..303Q} {506, 303}

\bibitem[\protect\citeauthoryear{{Rajpaul}, {Aigrain}, {Osborne}, {Reece}  \& {Roberts}}{{Rajpaul} et~al.}{2015}]{Rajpaul+2015}
{Rajpaul} V.,  {Aigrain} S.,  {Osborne} M.~A.,  {Reece} S.,   {Roberts} S.,  2015, \mn@doi [\mnras] {10.1093/mnras/stv1428}, \href {https://ui.adsabs.harvard.edu/abs/2015MNRAS.452.2269R} {452, 2269}

\bibitem[\protect\citeauthoryear{{Rajpaul}, {Aigrain}  \& {Roberts}}{{Rajpaul} et~al.}{2016}]{Rajpaul+2016}
{Rajpaul} V.,  {Aigrain} S.,   {Roberts} S.,  2016, \mn@doi [\mnras] {10.1093/mnrasl/slv164}, \href {https://ui.adsabs.harvard.edu/abs/2016MNRAS.456L...6R} {456, L6}

\bibitem[\protect\citeauthoryear{{Rescigno}, {Dixon}  \& {Haywood}}{{Rescigno} et~al.}{2023}]{Rescigno+2023}
{Rescigno} F.,  {Dixon} B.,   {Haywood} R.~D.,  2023, {MAGPy-RV: Gaussian Process regression pipeline with MCMC parameter searching}, Astrophysics Source Code Library, record ascl:2310.006 (\mn@eprint {ascl} {2310.006})

\bibitem[\protect\citeauthoryear{{Rescigno} et~al.,}{{Rescigno} et~al.}{2024a}]{Rescigno+2024}
{Rescigno} F.,  et~al., 2024a, \mn@doi [\mnras] {10.1093/mnras/stad3255}, \href {https://ui.adsabs.harvard.edu/abs/2024MNRAS.527.5385R} {527, 5385}

\bibitem[\protect\citeauthoryear{{Rescigno} et~al.,}{{Rescigno} et~al.}{2024b}]{Rescigno+2024b}
{Rescigno} F.,  et~al., 2024b, \mn@doi [\mnras] {10.1093/mnras/stae1634}, \href {https://ui.adsabs.harvard.edu/abs/2024MNRAS.532.2741R} {532, 2741}

\bibitem[\protect\citeauthoryear{{Ryabchikova}, {Piskunov}, {Kurucz}, {Stempels}, {Heiter}, {Pakhomov}  \& {Barklem}}{{Ryabchikova} et~al.}{2015}]{Ryabchikova+2015}
{Ryabchikova} T.,  {Piskunov} N.,  {Kurucz} R.~L.,  {Stempels} H.~C.,  {Heiter} U.,  {Pakhomov} Y.,   {Barklem} P.~S.,  2015, \mn@doi [\physscr] {10.1088/0031-8949/90/5/054005}, \href {https://ui.adsabs.harvard.edu/abs/2015PhyS...90e4005R} {90, 054005}

\bibitem[\protect\citeauthoryear{{Scargle}}{{Scargle}}{1982}]{Scargle1982}
{Scargle} J.~D.,  1982, \mn@doi [\apj] {10.1086/160554}, \href {https://ui.adsabs.harvard.edu/abs/1982ApJ...263..835S} {263, 835}

\bibitem[\protect\citeauthoryear{{Scherrer} et~al.,}{{Scherrer} et~al.}{2012}]{Scherrer+2012}
{Scherrer} P.~H.,  et~al., 2012, \mn@doi [Solar Physics] {10.1007/s11207-011-9834-2}, \href {https://ui.adsabs.harvard.edu/abs/2012SoPh..275..207S} {275, 207}

\bibitem[\protect\citeauthoryear{{Schou} et~al.,}{{Schou} et~al.}{2012}]{Schou+2012}
{Schou} J.,  et~al., 2012, \mn@doi [Solar Physics] {10.1007/s11207-011-9842-2}, \href {https://ui.adsabs.harvard.edu/abs/2012SoPh..275..229S} {275, 229}

\bibitem[\protect\citeauthoryear{{Serrano}, {Barros}, {Oshagh}, {Santos}, {Faria}, {Demangeon}, {Sousa}  \& {Lendl}}{{Serrano} et~al.}{2018}]{Serrano+2018}
{Serrano} L.~M.,  {Barros} S.~C.~C.,  {Oshagh} M.,  {Santos} N.~C.,  {Faria} J.~P.,  {Demangeon} O.,  {Sousa} S.~G.,   {Lendl} M.,  2018, \mn@doi [\aap] {10.1051/0004-6361/201731206}, \href {https://ui.adsabs.harvard.edu/abs/2018A&A...611A...8S} {611, A8}

\bibitem[\protect\citeauthoryear{{Valenti} \& {Fischer}}{{Valenti} \& {Fischer}}{2005}]{Valenti&Fischer2005}
{Valenti} J.~A.,  {Fischer} D.~A.,  2005, \mn@doi [\apjs] {10.1086/430500}, \href {https://ui.adsabs.harvard.edu/abs/2005ApJS..159..141V} {159, 141}

\bibitem[\protect\citeauthoryear{{Valenti} \& {Piskunov}}{{Valenti} \& {Piskunov}}{1996}]{Valenti&Piskunov1996}
{Valenti} J.~A.,  {Piskunov} N.,  1996, \aaps, \href {https://ui.adsabs.harvard.edu/abs/1996A&AS..118..595V} {118, 595}

\bibitem[\protect\citeauthoryear{{Wehrhahn}, {Piskunov}  \& {Ryabchikova}}{{Wehrhahn} et~al.}{2023}]{Wehrhahn+2023}
{Wehrhahn} A.,  {Piskunov} N.,   {Ryabchikova} T.,  2023, \mn@doi [\aap] {10.1051/0004-6361/202244482}, \href {https://ui.adsabs.harvard.edu/abs/2023A&A...671A.171W} {671, A171}

\bibitem[\protect\citeauthoryear{{Wilson}}{{Wilson}}{1968}]{Wilson1968}
{Wilson} O.~C.,  1968, \mn@doi [\apj] {10.1086/149652}, \href {https://ui.adsabs.harvard.edu/abs/1968ApJ...153..221W} {153, 221}

\bibitem[\protect\citeauthoryear{{Zechmeister} \& {K{\"u}rster}}{{Zechmeister} \& {K{\"u}rster}}{2009}]{Zechmeister&Kurster2009}
{Zechmeister} M.,  {K{\"u}rster} M.,  2009, \mn@doi [\aap] {10.1051/0004-6361:200811296}, \href {https://ui.adsabs.harvard.edu/abs/2009A&A...496..577Z} {496, 577}

\makeatother
\end{thebibliography}








\bsp	
\label{lastpage}
\end{document}